\documentclass[12pt,letterpaper]{article}
\usepackage[a4paper, total={7in, 10in}]{geometry}

\usepackage{graphicx}
\usepackage{helvet}
\usepackage{authblk}
\usepackage{hyperref}
\usepackage{amsmath} 
\usepackage{amssymb} 
\usepackage{orcidlink} 
\usepackage{booktabs}  
\usepackage{placeins}  
\usepackage{soul}
\usepackage[super,comma,sort&compress]  
   {natbib}
\usepackage[right]{lineno} \linenumbers

\makeatletter
\renewcommand{\maketitle}{\bgroup\setlength{\parindent}{0pt}
\begin{flushleft}
  \textbf{\@title}
  
  \@author
\end{flushleft}\egroup}
\makeatother

\title{Proteome-wide prediction of mode of inheritance and molecular mechanism underlying genetic diseases using structural interactomics}
\date{}

\author[1,2]{Ali Saadat}
\author[1,2,3,*]{Jacques Fellay}

\affil[1]{School of Life Sciences, Ecole Polytechnique Fédérale de Lausanne, Lausanne, Switzerland}
\affil[2]{Swiss Institute of Bioinformatics, Lausanne, Switzerland}
\affil[3]{Precision Medicine Unit, Biomedical Data Science Center, Lausanne University Hospital and University of Lausanne, Lausanne, Switzerland}

\affil[*]{Correspondence: jacques.fellay@epfl.ch}

\begin{document}

\maketitle

\section*{SUMMARY}

Genetic diseases can be classified according to their modes of inheritance and their underlying molecular mechanisms. Autosomal dominant disorders often result from DNA variants that cause loss-of-function, gain-of-function, or dominant-negative effects, while autosomal recessive diseases are primarily linked to loss-of-function variants. In this study, we introduce a graph-of-graphs approach that leverages protein-protein interaction networks and high-resolution protein structures to predict the mode of inheritance of diseases caused by variants in autosomal genes, and to classify dominant-associated proteins based on their functional effect. Our approach integrates graph neural networks, structural interactomics and topological network features to provide proteome-wide predictions, thus offering a scalable method for understanding genetic disease mechanisms.

\section*{KEYWORDS}


Graph neural networks, protein structure, mode of inheritance, graph-of-graphs, protein-protein interaction, genetic diseases

\section*{INTRODUCTION}

Most human genetic diseases result from variants that disrupt protein function through diverse molecular mechanisms, which play a critical role in determining their mode of inheritance (MOI) \citep{Zschocke2023}. In autosomal dominant (AD) disorders, a single copy of a mutated gene can result in disease, often through loss of function (LOF) due to haploinsufficiency (HI), where the remaining wild-type allele cannot compensate for the lost function \citep{Veitia2002-mz}. Dominant disorders can also result from non-LOF mechanisms, such as gain of function (GOF), where the mutant protein acquires a new or altered function, and the dominant-negative (DN) effect, where the mutant protein interferes with the normal function of the wild-type protein \citep{Backwell2022-iu}. In contrast, autosomal recessive (AR) disorders require variants in both gene copies, predominantly involving LOF mechanisms, such as missense variants that destabilize protein structure or nonsense variants leading to truncated, non-functional proteins. 

Previous studies on MOI prediction have introduced computational tools such as DOMINO \citep{Quinodoz2017-sc}, which utilizes linear discriminant analysis (LDA) to predict whether a protein is associated with AD disorders by integrating various features such as genomic data, conservation, and protein interactions. MOI-Pred \citep{Petrazzini2024-ue}, on the other hand, focuses on variant-level predictions, specifically targeting missense variants associated with AR diseases.

More recent research has aimed at predicting the functional impact of variants in specific genes. LoGoFunc combines gene-, protein-, and variant-level features to predict pathogenic GOF, LOF, and neutral variants \citep{Stein2023-dh}. Another study explored the structural effects of variants, finding that non-LOF variants tend to have milder impacts on protein structure \citep{Gerasimavicius2022-rx}. Additionally, a recent study employed three support vector machines (SVM) to predict protein coding genes associated with DN, GOF, and HI mechanisms \citep{Badonyi2024}.

\begin{figure}[h]
    \centering
    \includegraphics[width=0.95\textwidth]{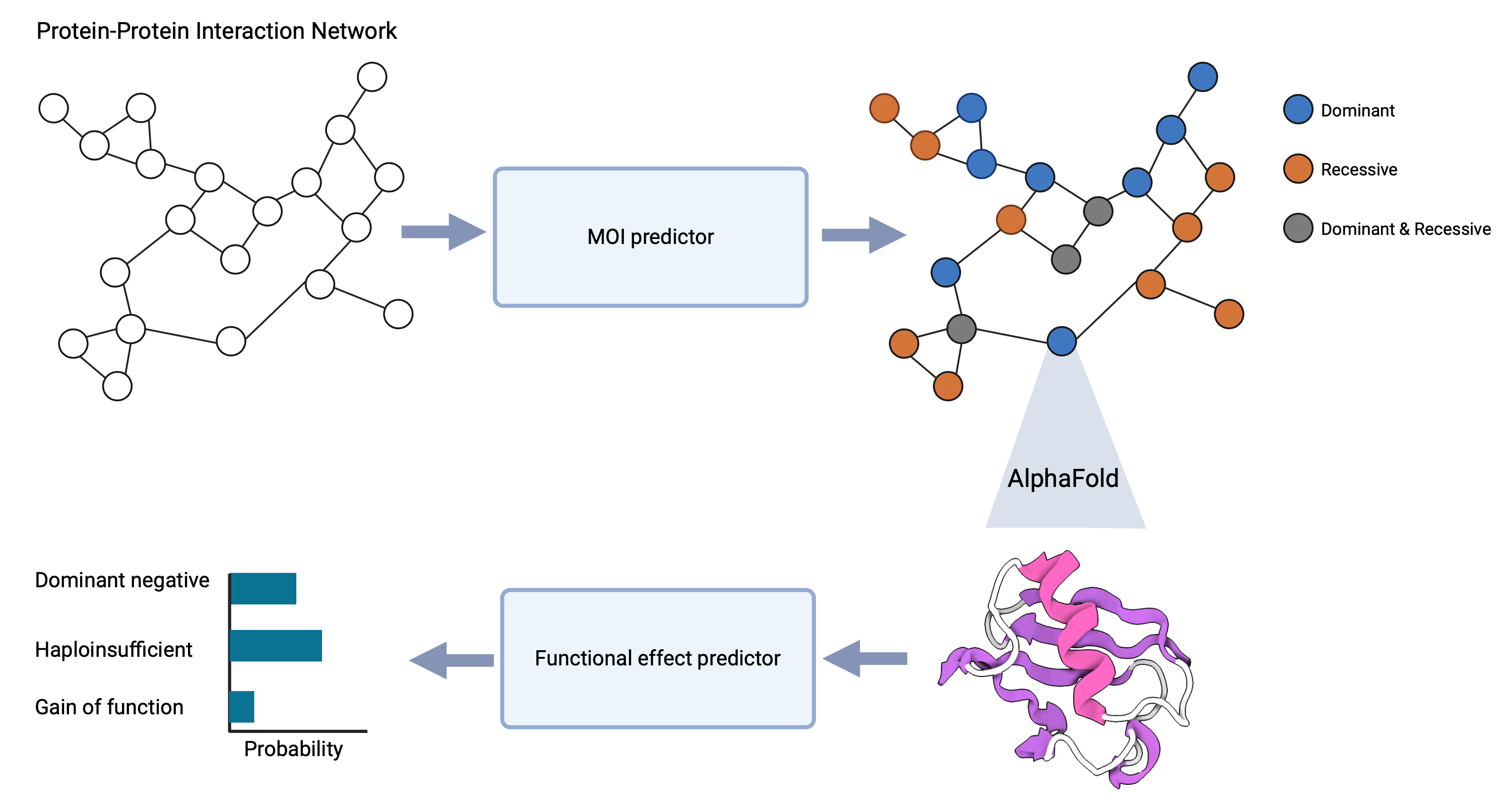} 
    \caption{\footnotesize Overview of the study: at first the mode of inheritance (MOI) is predicted for all of the autosomal proteins in the protein-protein interaction network. Afterwards, AlphaFold protein structures are used to generate residue graphs for each dominant protein, and functional effects are predicted based on these graphs. Figure created with \url{BioRender.com}.}
    \label{overview}
\end{figure}

In this study, we present a comprehensive approach for predicting the MOI for all proteins encoded by autosomal genes, as well as elucidating the functional effect of variants underlying AD genetic disorders (Figure \ref{overview}). Our framework combines graph neural networks (GNNs) \citep{zhou2021graphneuralnetworksreview} with structural interactomics by creating a graph-of-graphs \citep{DAgostino2014-tm}, utilizing both protein-protein interaction (PPI) network and high-resolution protein structures. For MOI prediction, we model proteins as nodes within the PPI network, incorporating topological and protein-level features for classification. For molecular mechanism prediction, we represent each protein as a graph of amino acid residues, leveraging structure-based features to classify the functional effect as HI, GOF, or DN. This integrated approach enables proteome-wide prediction of inheritance patterns and provides mechanistic insights into AD diseases, offering a novel, scalable framework for understanding genetic disorders.

For the sake of flow and conciseness, we refer to "proteins associated with AD disorders" as AD proteins and "proteins associated with AR disorders" as AR proteins. Similarly, we use DN (GOF/LOF) proteins instead of "proteins associated with DN (GOF/LOF) molecular disease mechanisms".

\section*{RESULTS}

\label{results}

\subsection*{Datasets}

\paragraph{MOI data}

\begin{figure}[h]
    \centering
    \begin{minipage}{0.34\textwidth}
        \centering
        \includegraphics[width=\textwidth]{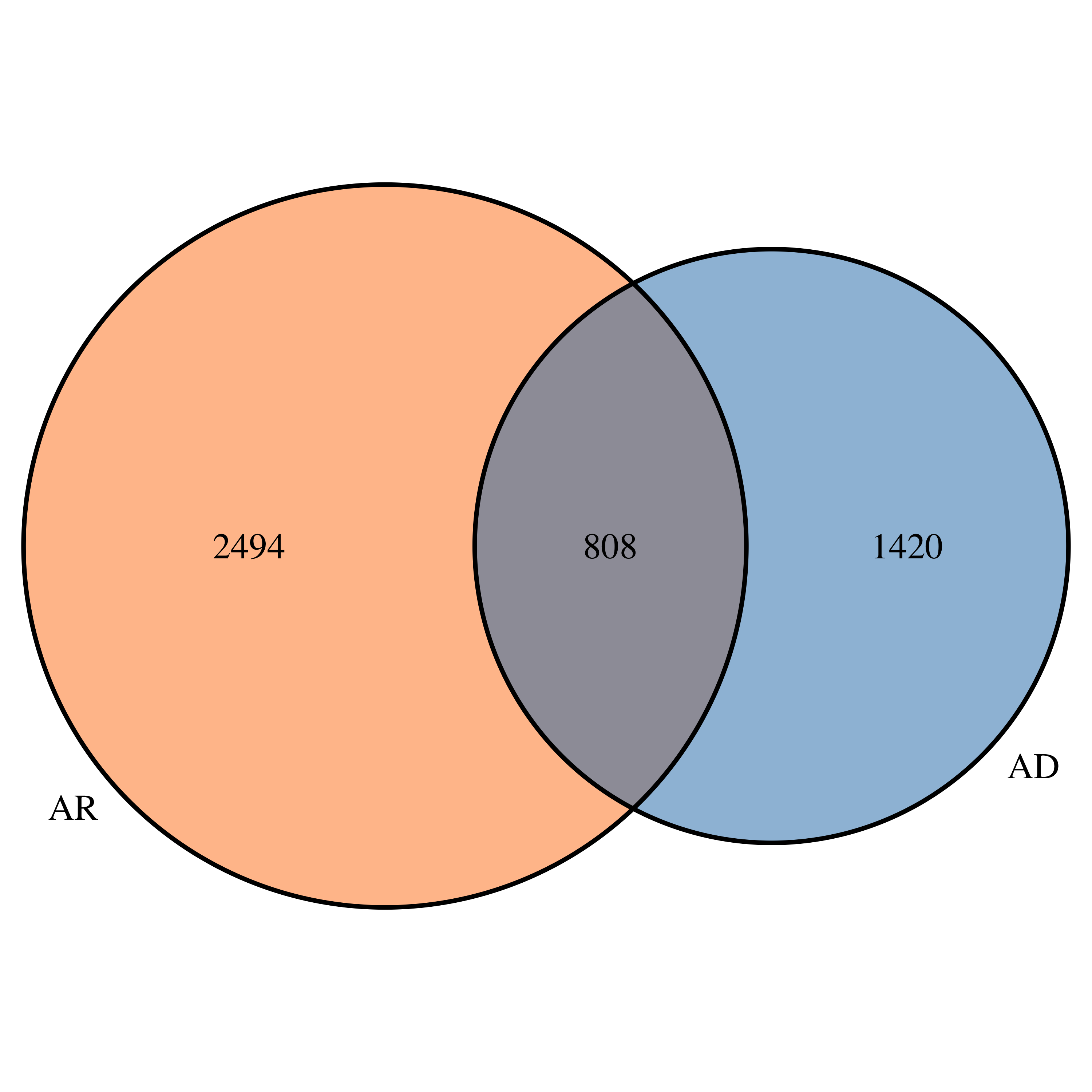}
    \end{minipage}
    \hspace{0.1\textwidth}
    \begin{minipage}{0.34\textwidth}
        \centering
        \includegraphics[width=\textwidth]{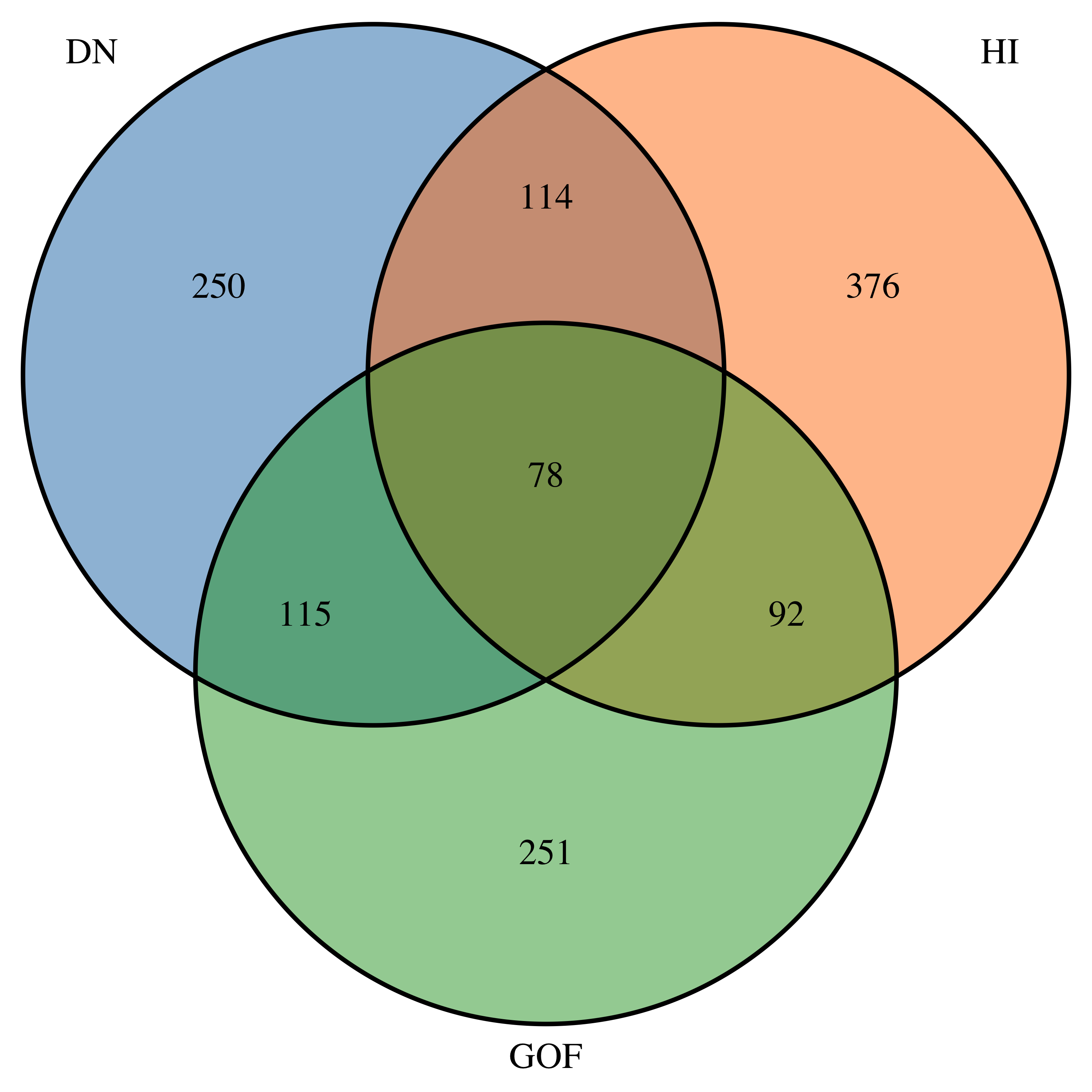}
    \end{minipage}
    \caption{\footnotesize The number of proteins with labeled MOI (left) and molecular mechanism (right).}
    \label{venn_diagram_ADAR_DNLOFGOF}
\end{figure}

We gathered 4,737 MOI-labeled proteins, among them 2,494 (53\%) were only AR, 1,420 (30\%) were only AD, and 808 (17\%) were both AD and AR (Figure \ref{venn_diagram_ADAR_DNLOFGOF}, left). 

\paragraph{Functional effect data} 

We collected 1,276 proteins with annotated functional effect, among them 250 (20\%) were only DN, 376 (29\%) were only HI, 251 (20\%) were only GOF, 114 (9\%) were both DN and HI, 115 (9\%) were both DN and GOF, 92 (7\%) were both HI and GOF, and 78 (6\%) were all of the DN, HI, GOF (Figure \ref{venn_diagram_ADAR_DNLOFGOF}, right).

\paragraph{PPI construction and annotation}  
We constructed a comprehensive PPI network comprising 17,248 nodes and 375,494 edges by integrating interactions from STRINGdb \citep{Szklarczyk2022}, BioGRID \citep{Oughtred2020}, the Human Reference Interactome (HuRI) \citep{Luck2020}, and \citet{Menche2015}. To characterize proteins, we annotated them with 78 selected features covering structural, functional, evolutionary, and regulatory properties (Supplementary Table 1).  

\paragraph{Protein graph construction and annotation}  
We obtained predicted protein structures from the AlphaFold database \citep{Varadi2023} and constructed residue-level graphs using Graphein \citep{jamasb2022graphein}. In these graphs, nodes represent amino acids, while edges capture peptide bonds, hydrogen bonds, disulfide bonds, ionic interactions, and other structural contacts, including long-range interactions. We annotated amino acid residues with 73 selected features reflecting structural, sequence-based, biochemical, and evolutionary characteristics (Supplementary Table 2).

\subsection*{Model development}

\paragraph{Study design}  

We formulated MOI prediction as a node classification task within the PPI network and functional effect prediction as a graph classification task. Both models employed a multi-label classification approach, allowing each input to have multiple labels. We evaluated various graph neural network architectures, including graph convolutional networks (GCN) \citep{kipf2017}, graph attention networks (GAT) \citep{GATv2}, and graph isomorphism networks (GIN) \citep{xu2019}.

\paragraph{Data splitting}  

To create training, validation, and test sets, we clustered protein sequences using MMseqs2 \citep{Steinegger2017} with thresholds of 20\% coverage and 20\% sequence identity. Proteins were then split into 80\% training, 10\% validation, and 10\% test sets.

\paragraph{Hyperparameter tuning and model training}  

All models used a single hidden layer, with the output layer containing two units for MOI prediction (AD and AR) and three units for functional effect prediction (DN, HI, and GOF). To determine the optimal configurations, we evaluated 25 hyperparameter combinations on the validation set, varying the hidden layer size across five values (128, 64, 32, 16, and 8) and the learning rate across five values ranging from \(10^{-2}\) to \(5 \times 10^{-4}\). The results of hyperparameter tuning for MOI and functional effect prediction are provided in Supplementary Tables 3 and 4. Using the selected hyperparameters, we trained each model with binary cross-entropy loss for up to 100 epochs, applying early stopping based on validation loss to prevent overfitting.

\subsection*{Models performance evaluation}

\paragraph{MOI models}

We evaluated all trained models on the unseen test set (Table \ref{MOI-comparison-table}). The GCN model achieved the highest precision score, while the GAT model had the best recall and $F_1$ score. Due to the class imbalance in the MOI dataset, we prioritized maximizing $F_1$ score and selected the GAT model. We also assessed the performance of DOMINO \citep{Quinodoz2017-sc} as outlined in the methods section, and found that our models outperformed it (Table \ref{MOI-comparison-table}).

\FloatBarrier
\begin{table}
  \caption{MOI prediction performance on the test set}
  \label{MOI-comparison-table}
  \centering
  \begin{tabular}{lcccc}
    \toprule
    Metric         & GCN    & GAT    & GIN  & LDA \citep{Quinodoz2017-sc}  \\
    \midrule
    F1        & 0.745 & \textbf{0.750}  & 0.671 & 0.685  \\
    Precision  & \textbf{0.776}  & 0.770  & 0.764 & 0.721 \\
    Recall    & 0.725 & \textbf{0.731}  & 0.621 & 0.654 \\
    \bottomrule
  \end{tabular}
\end{table}

\paragraph{Functional effect models}

Table \ref{mechanism-comparison-table} shows the performance of various models on the functional effect test set, with the GCN model achieving the highest $F_1$ score. We also evaluated the SVM models from \citet{Badonyi2024} as described in the methods section. Based on the overall performance, we selected the GCN model for functional effect prediction.

\begin{table}
  \caption{Functional effect prediction performance on the test set}
  \label{mechanism-comparison-table}
  \centering
  \begin{tabular}{lccccc}
    \toprule
    Metric         & GCN    & GAT    & GIN  &  SVM \citep{Badonyi2024} \\
    \midrule
    F1        & \textbf{0.627}  & 0.590  & 0.600 & 0.593  \\
    Precision  & 0.605  & 0.517  & 0.549 & \textbf{0.669} \\
    Recall    & 0.659  & \textbf{0.712}  & 0.676  & 0.535\\
    \bottomrule
  \end{tabular}
\end{table}

\subsection*{Models interpretation}

\paragraph{MOI feature attribution}

\begin{figure}[h]
    \centering
    \begin{minipage}{0.45\textwidth}
        \centering
        \includegraphics[width=\textwidth]{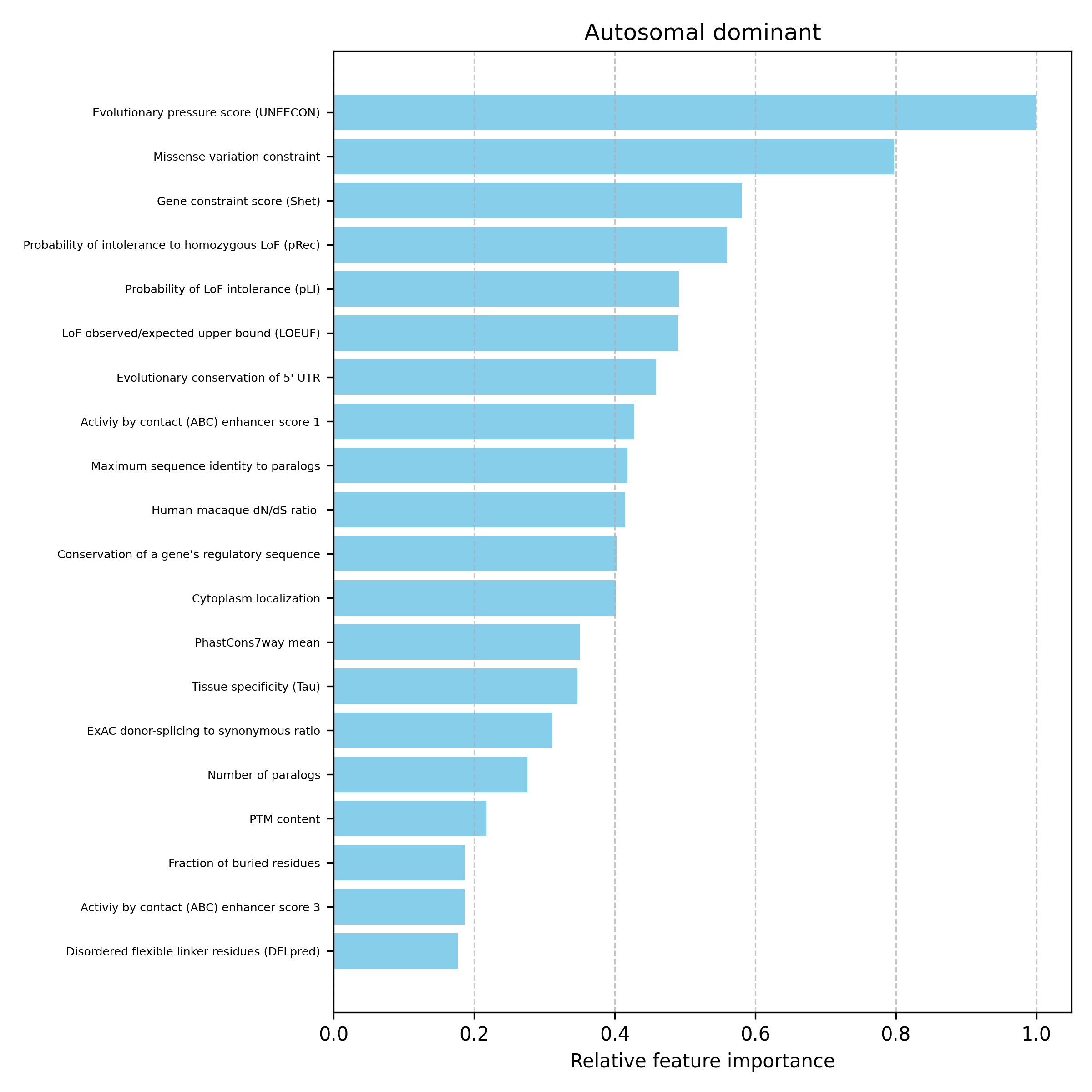}
    \end{minipage}
    \hspace{0.05\textwidth}
    \begin{minipage}{0.43\textwidth}
        \centering
        \includegraphics[width=\textwidth]{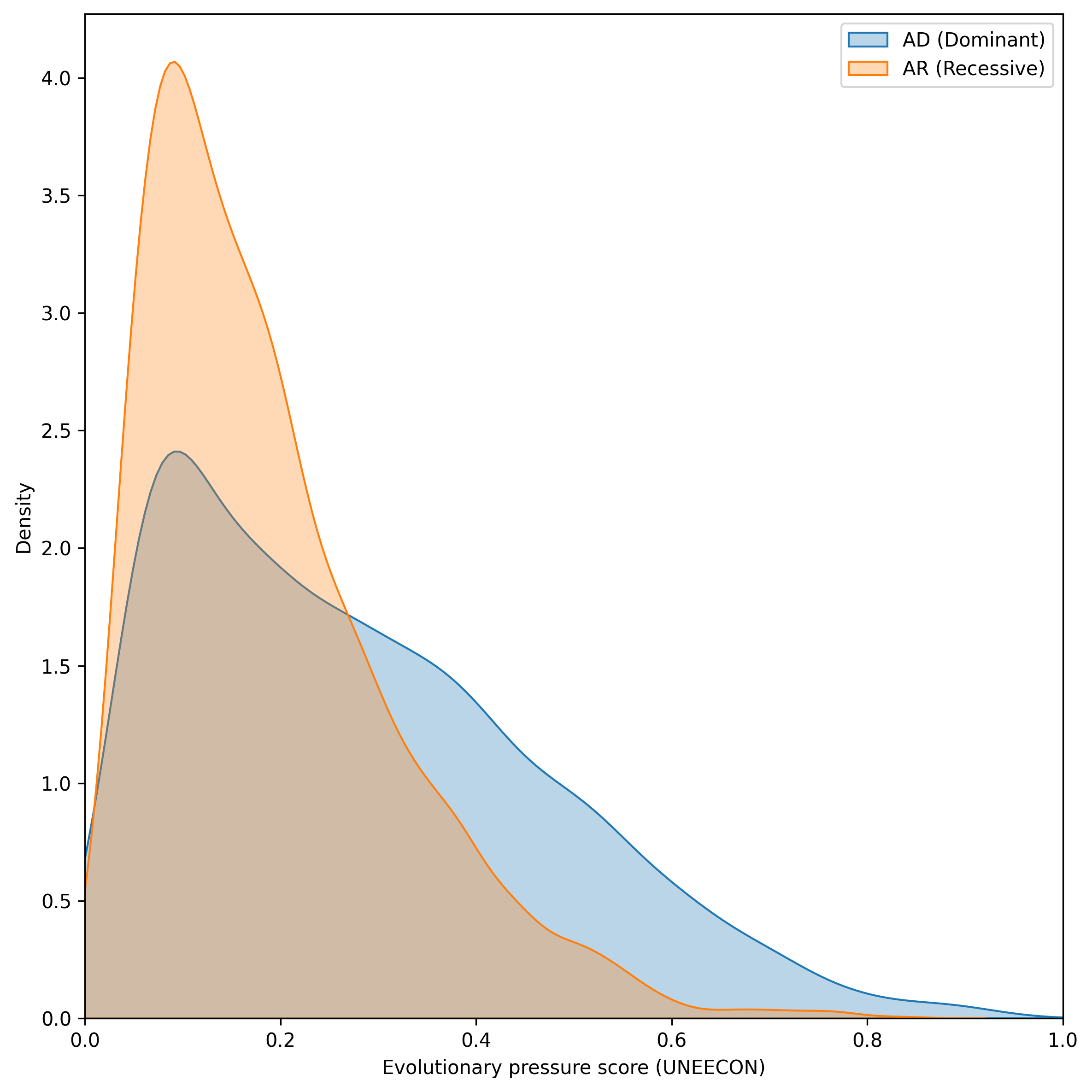}
    \end{minipage}
    \caption{\footnotesize GAT model interpretation for AD prediction (left) and UNEECON score distribution for AD and AR genes (right).}
    \label{AD_featuure_importance}
\end{figure}

Using the GAT model, we calculated features attribution separately for correctly predicted AD or AR proteins in the test set. We observed that the most important predictors for AD prediction are features related to constraint and conservation (Figure \ref{AD_featuure_importance} left). The top feature was UNEECON, which measures the evolutionary pressure \citep{Huang2020-pd}. Using the labeled data, we observed that AD proteins have higher UNEECON values compared to AR proteins (Figure \ref{AD_featuure_importance} right). 

\begin{figure}[h]
    \centering
    \begin{minipage}{0.45\textwidth}
        \centering
        \includegraphics[width=\textwidth]{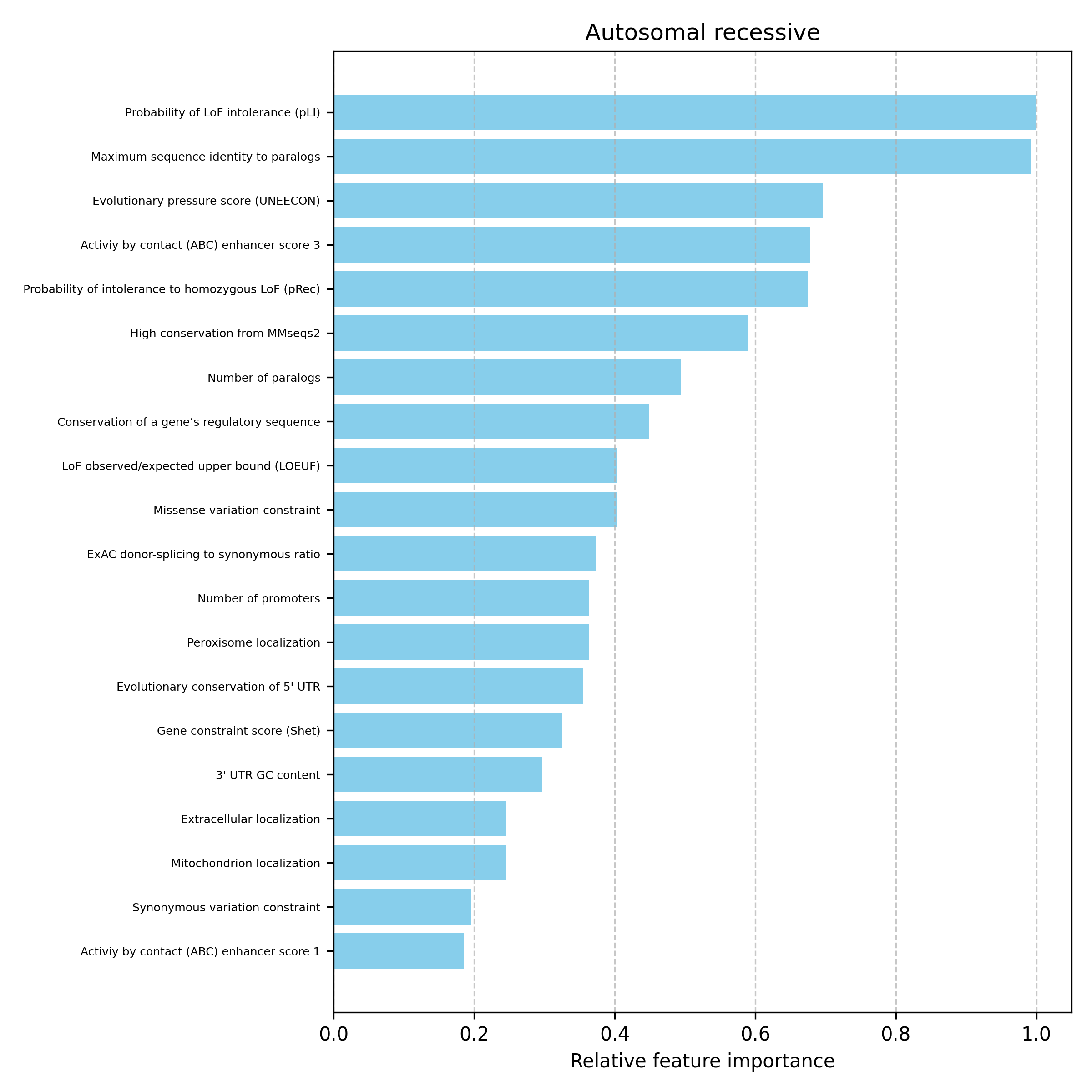}
    \end{minipage}
    \hspace{0.05\textwidth}
    \begin{minipage}{0.43\textwidth}
        \centering
        \includegraphics[width=\textwidth]{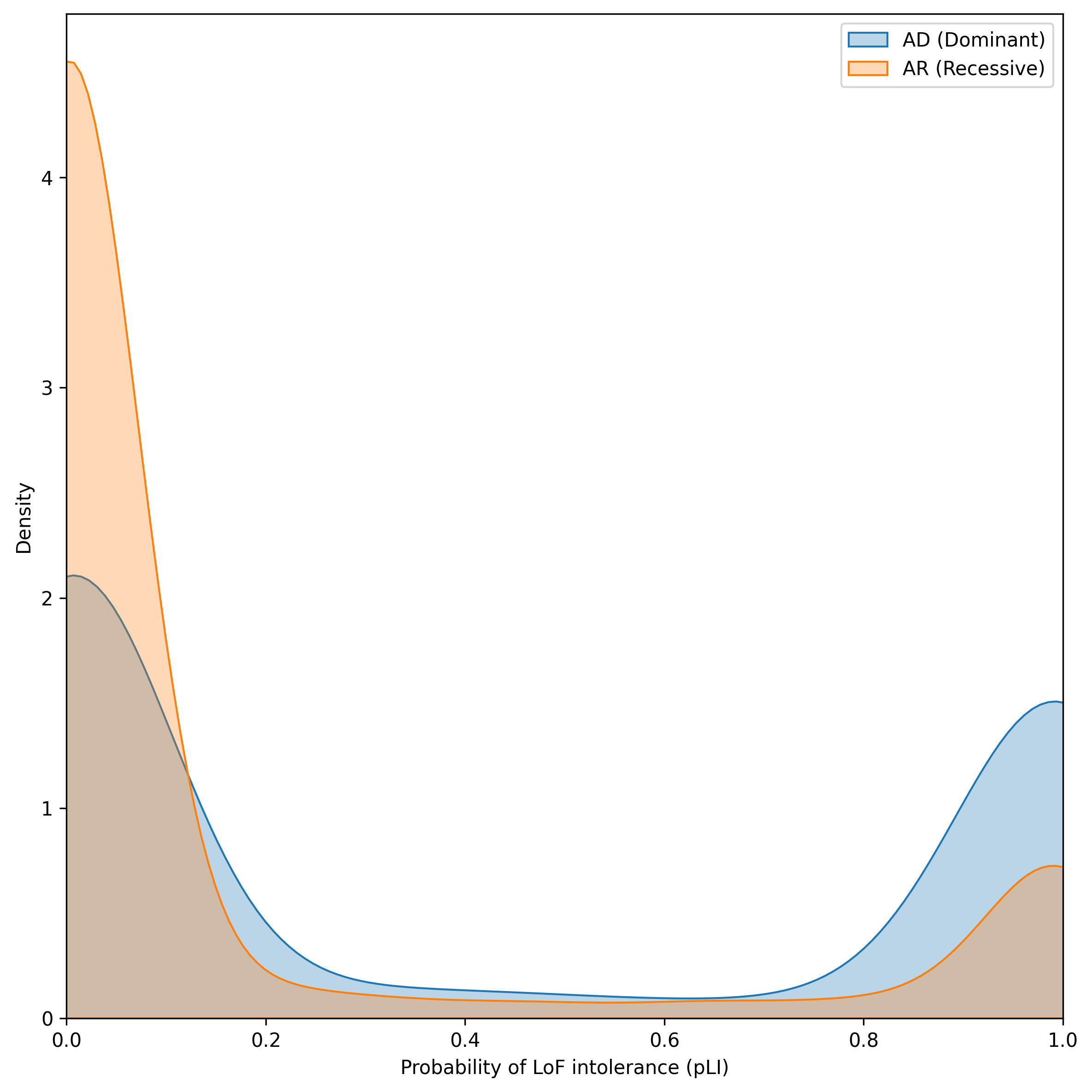}
    \end{minipage}
    \caption{\footnotesize GAT model interpretation for AR prediction (left) and pLI distribution for AD and AR genes (right).}
    \label{AR_featuure_importance}
\end{figure}

For AR prediction, the most important feature was pLI, which is probability of loss-of-function intolerance \citep{exac2016} (Figure \ref{AR_featuure_importance} left). Using the ground truth dataset, we observed that AR proteins have lower pLI values compared to AD proteins (Figure \ref{AR_featuure_importance} right).

\paragraph{Functional effect feature attribution}

Using the GCN model, we measured features attribution for correctly predicted DN, HI, and GOF proteins. Because features are at residue-level and prediction are at protein-level, we cannot draw direct conclusions from these measurements, yet they can help to understand the associations. 

\begin{figure}[h]
    \centering
    \begin{minipage}{0.3\textwidth}
        \centering
        \includegraphics[width=\textwidth]{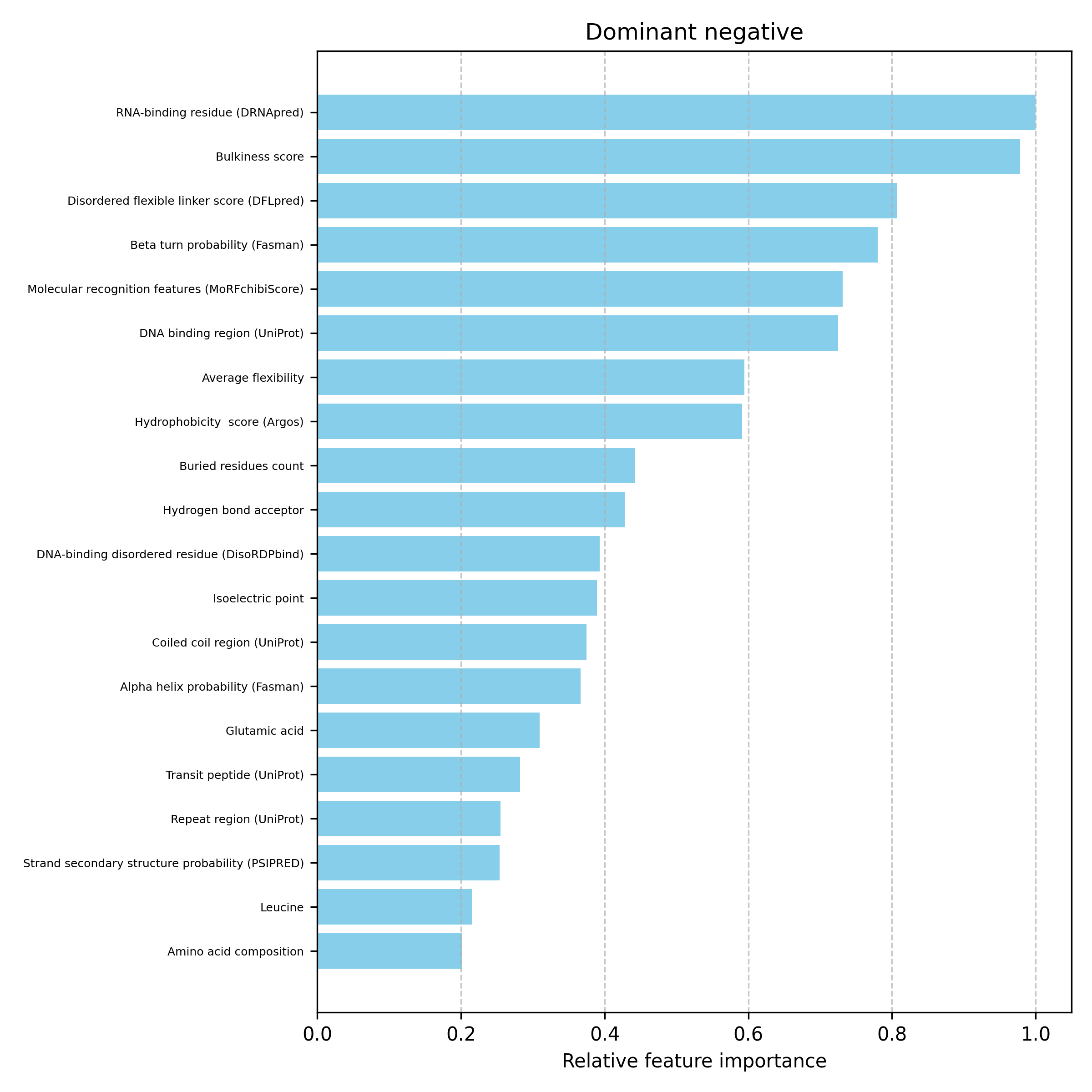}
    \end{minipage}
    \hspace{0.03\textwidth}
    \begin{minipage}{0.3\textwidth}
        \centering
        \includegraphics[width=\textwidth]{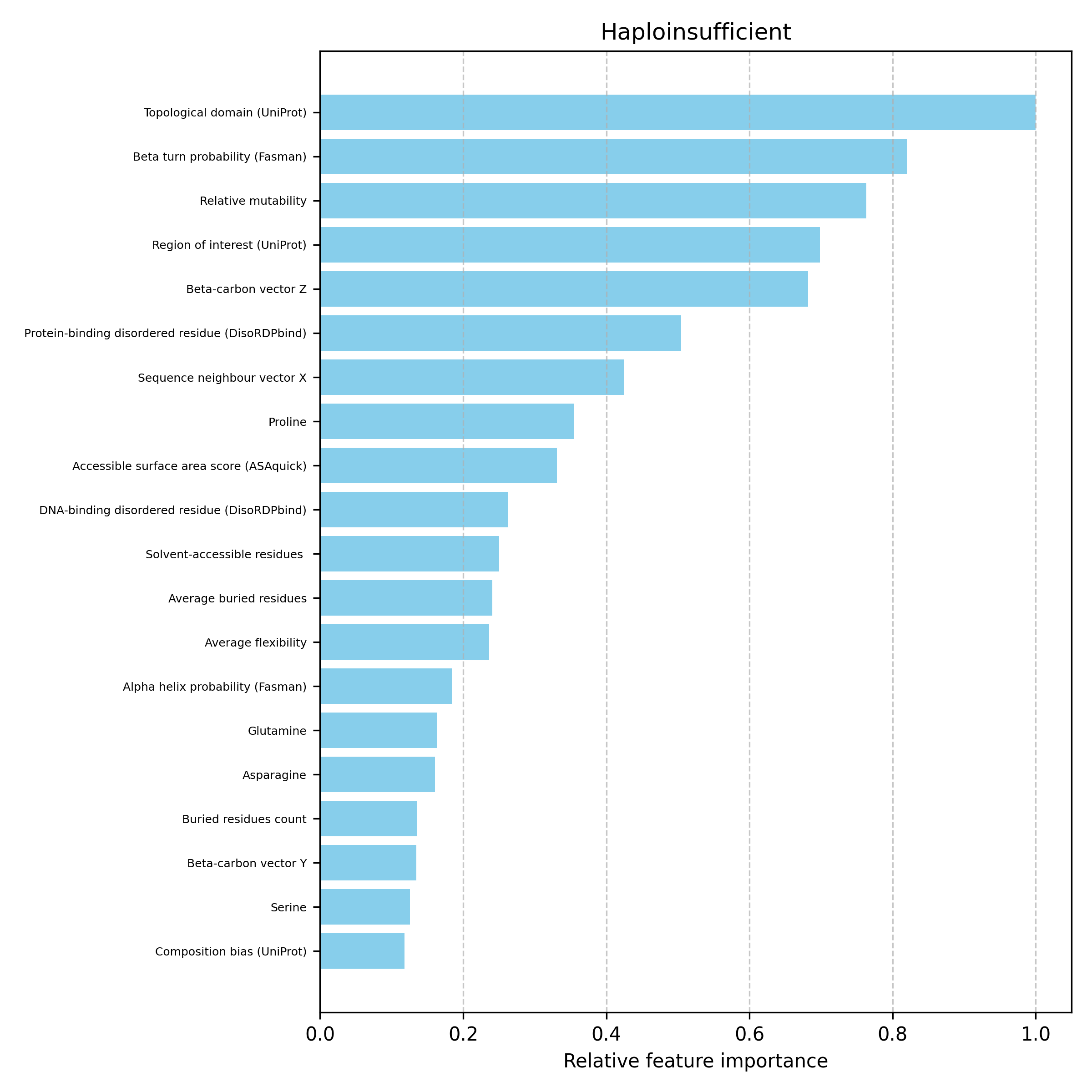}
    \end{minipage}
    \hspace{0.03\textwidth}
    \begin{minipage}{0.3\textwidth}
        \centering
        \includegraphics[width=\textwidth]{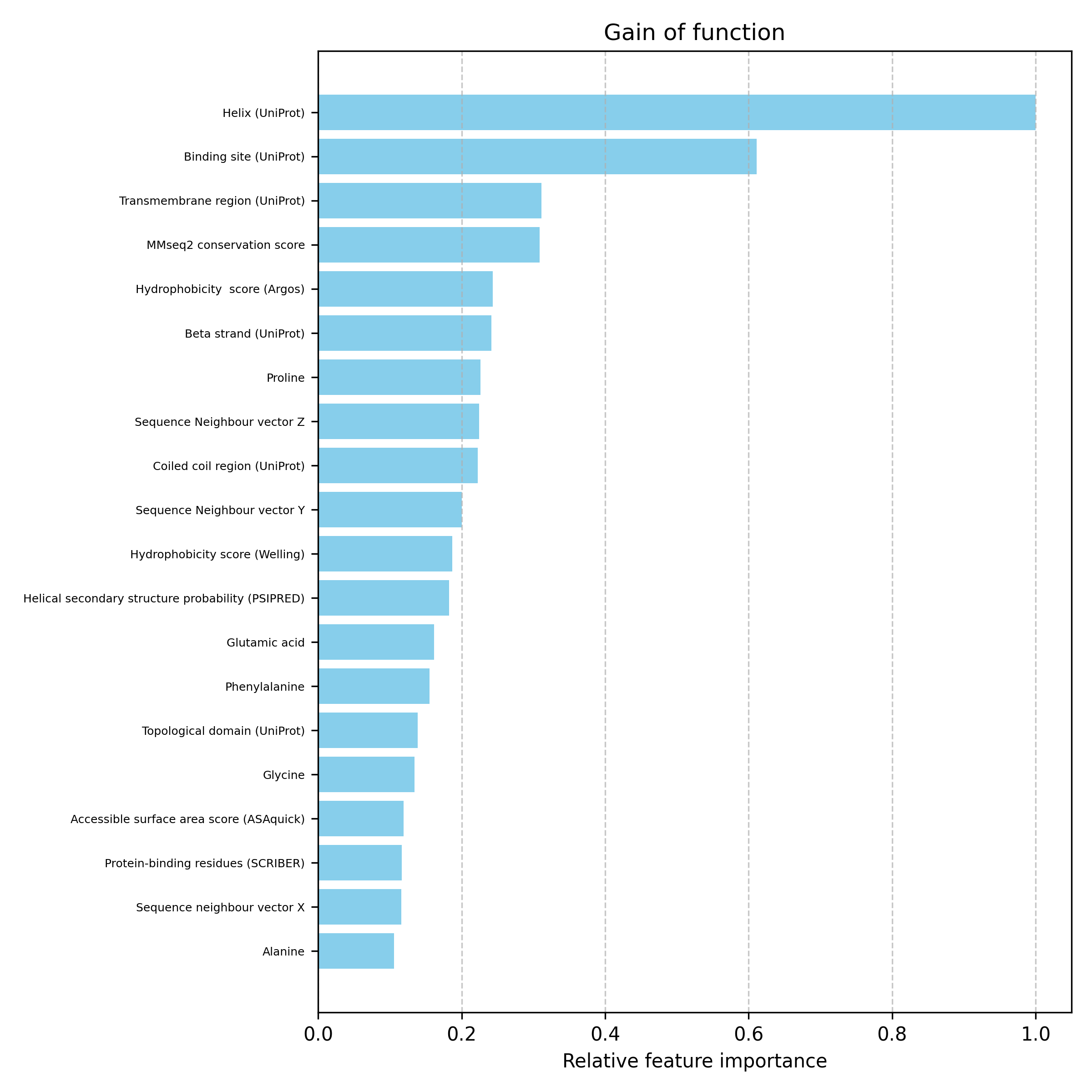}
    \end{minipage}
    \caption{\footnotesize GCN model interpretation for DN (left), HI (middle), and GOF (right) predictions.}
    \label{DN_HI_GOF_feature_importance}
\end{figure}

For DN proteins, the most important feature was the RNA-binding score based on DRNApred \citep{Yan2017-zi} (Figure~\ref{DN_HI_GOF_feature_importance}, left). Using the labeled data, we observed that residues in DN proteins have higher RNA-binding scores compared to HI and GOF proteins (Supplementary Figure S1).

For HI proteins, as shown in Figure~\ref{DN_HI_GOF_feature_importance} (middle), topological domain is the strongest predictor. This feature was derived from UniProt \citep{UniProt2022}. We observed that HI proteins have a lower fraction of topological domains compared to DN and GOF proteins (Supplementary Figure S2).

Feature attribution analysis for GOF proteins showed that the top feature is the helix structure (Figure~\ref{DN_HI_GOF_feature_importance}, right), derived from UniProt \citep{UniProt2022}. The distribution of helical fractions indicates that GOF proteins have a relatively higher fraction of helical structures compared to HI and DN proteins (Supplementary Figure S3).

\subsection*{Proteome-wide inference}

\paragraph{MOI prediction for all autosomal proteins}

Of the 17,248 nodes in the PPI network, 16,477 (96\%) were autosomal, and we used the GAT model to predict the most likely MOI for all of them. A total of 8,869 (54\%) were predicted to be AR, 6,277 (38\%) were predicted to be AD, and 1,206 (7\%) were predicted to be ADAR (Supplementary Figure S4). As expected, we observed a strong negative correlation between the probability of being AD and AR (Pearson correlation coefficient = -0.95) (Supplementary Figure S5).  

\begin{figure}[h]
    \centering
    \begin{minipage}{0.3\textwidth}
        \centering
        \includegraphics[width=\textwidth]{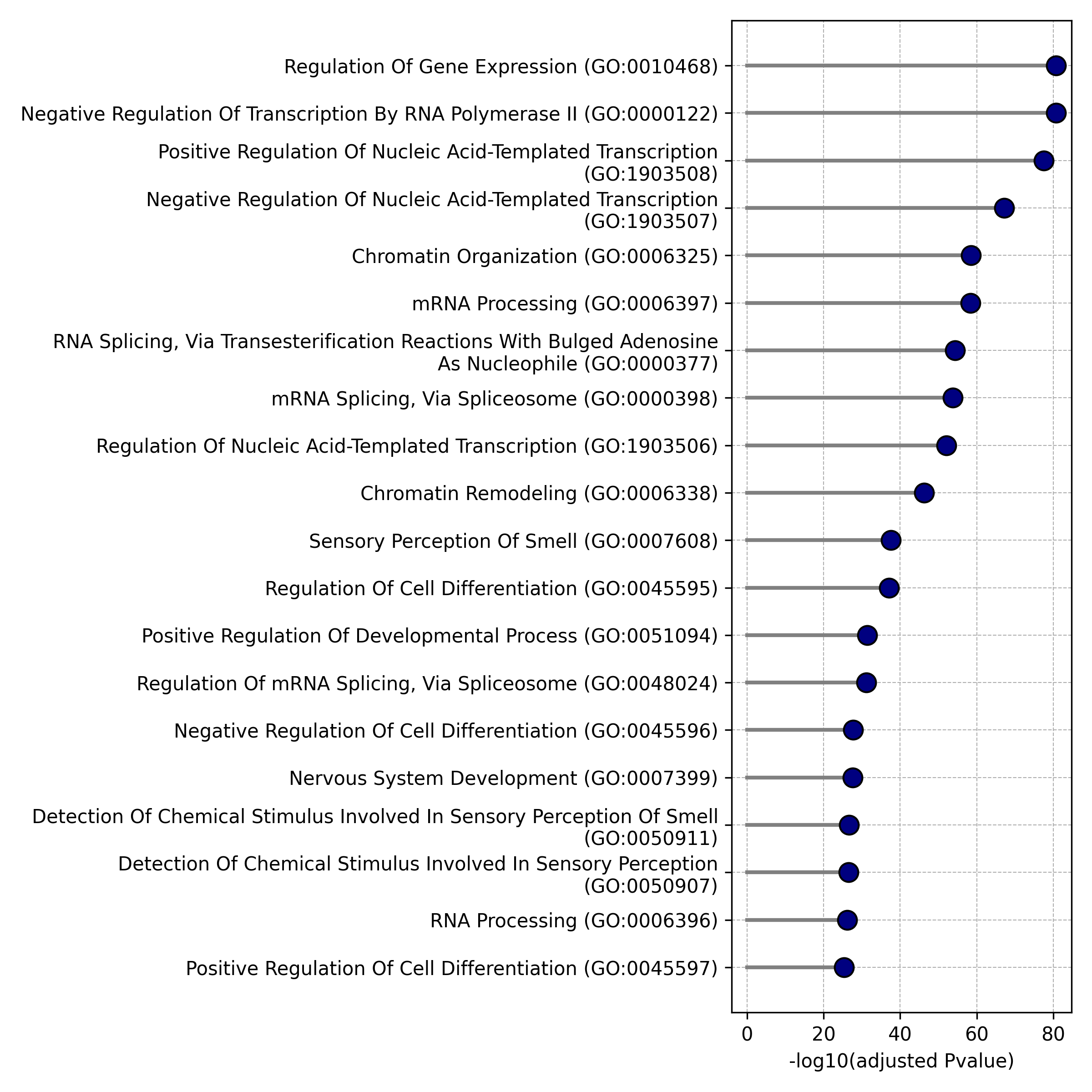}
    \end{minipage}
    \hspace{0.03\textwidth}
    \begin{minipage}{0.3\textwidth}
        \centering
        \includegraphics[width=\textwidth]{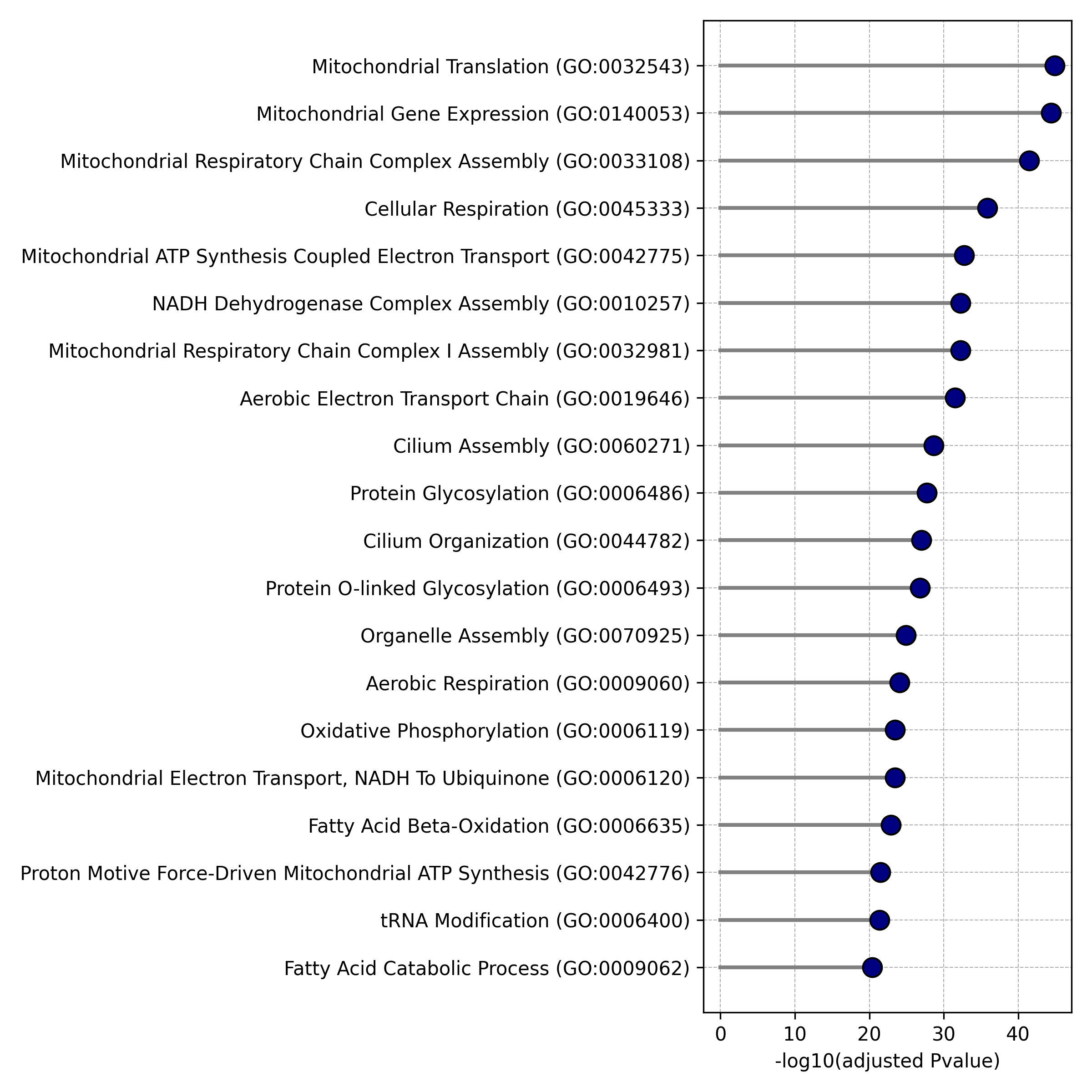}
    \end{minipage}
    \hspace{0.03\textwidth}
    \begin{minipage}{0.3\textwidth}
        \centering
        \includegraphics[width=\textwidth]{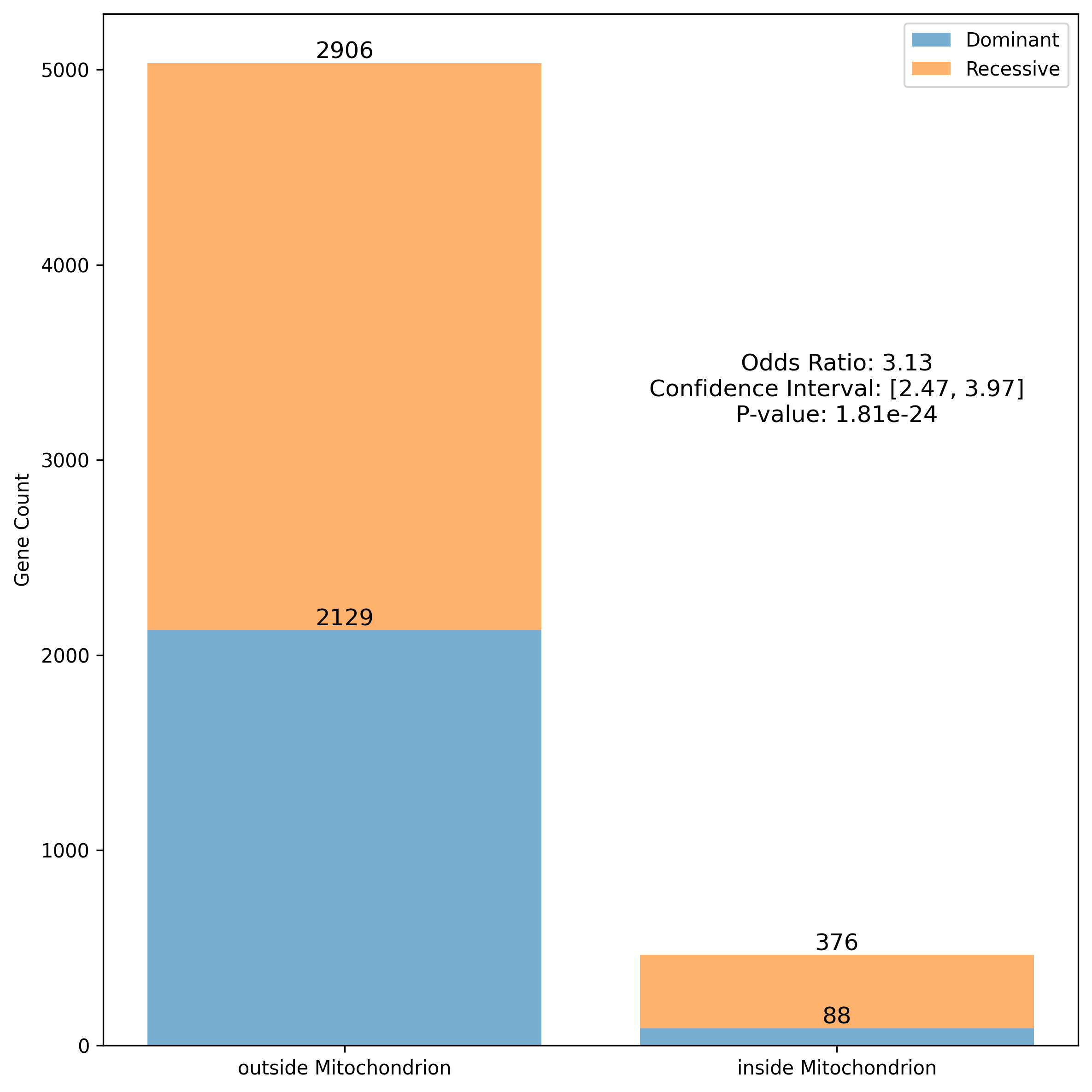}
    \end{minipage}
    \caption{\footnotesize Pahtway enrichment analysis for AD (left) and AR (middle) proteins. Right panel shows the number of proteins associated with sub-cellular localization inside or outside mitochondria. The odds ratio was calculated as 
    $\left(\frac{\text{AR\_inside}}{\text{AR\_outside}}\right) \Big/ \left(\frac{\text{AD\_inside}}{\text{AD\_outside}}\right)$. P-value was calculated using the Fisher's exact test.}
    \label{ADAR_pathways}
\end{figure}

Finally, we performed pathway enrichment analyses for AD and AR proteins separately. AD proteins were significantly enriched in pathways associated with gene regulation (Figure~\ref{ADAR_pathways}, left), while AR proteins were significantly overrepresented in mitochondrial pathways (Figure~\ref{ADAR_pathways}, middle). Using the ground truth dataset, we observed that AR proteins are more likely to be localized inside mitochondria compared to AD proteins (\(OR = 3.13, CI = [2.47, 3.97]\)) (Figure~\ref{ADAR_pathways}, right).

\paragraph{Functional effect prediction for all AD-predicted proteins}

Based on the proteome-wide MOI predictions, we identified 7,483 AD or ADAR proteins and predicted their functional effect using the GCN model. Among them, 2,043 (28\%) were classified as only DN, 1,097 (15\%) as only HI, and 415 (6\%) as only GOF. Additionally, 1,843 (26\%) were both DN and HI, 1,569 (22\%) were both DN and GOF, 181 (3\%) were both HI and GOF, and 35 (1\%) were classified as DN, HI, and GOF (Supplementary Figure S6). We also provide the counts based on AD-only and ADAR-only proteins in Supplementary Figures S7 and S8, respectively.

\begin{figure}[h]
    \centering
    \begin{minipage}{0.3\textwidth}
        \centering
        \includegraphics[width=\textwidth]{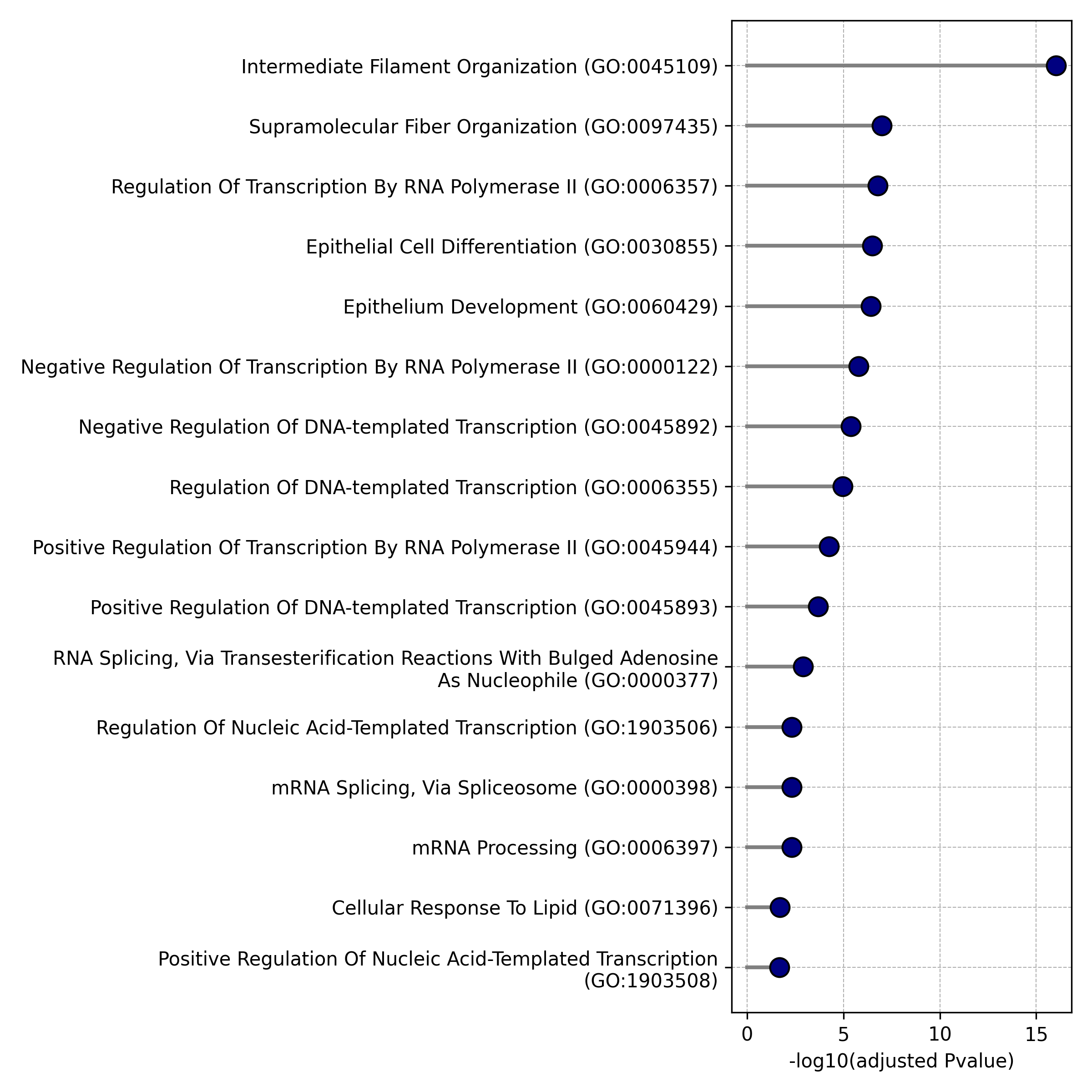}
    \end{minipage}
    \hspace{0.03\textwidth}
    \begin{minipage}{0.3\textwidth}
        \centering
        \includegraphics[width=\textwidth]{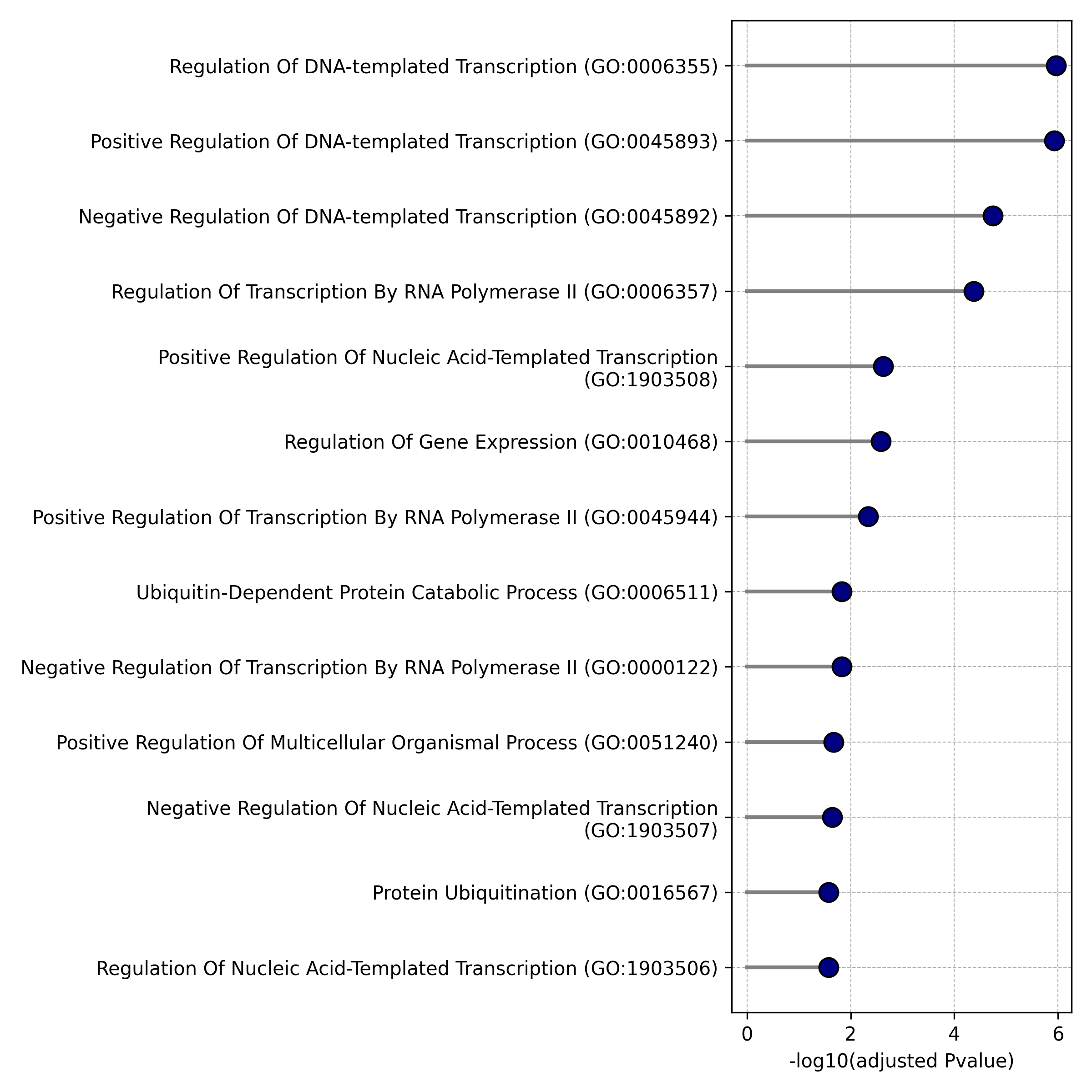}
    \end{minipage}
    \hspace{0.03\textwidth}
    \begin{minipage}{0.3\textwidth}
        \centering
        \includegraphics[width=\textwidth]{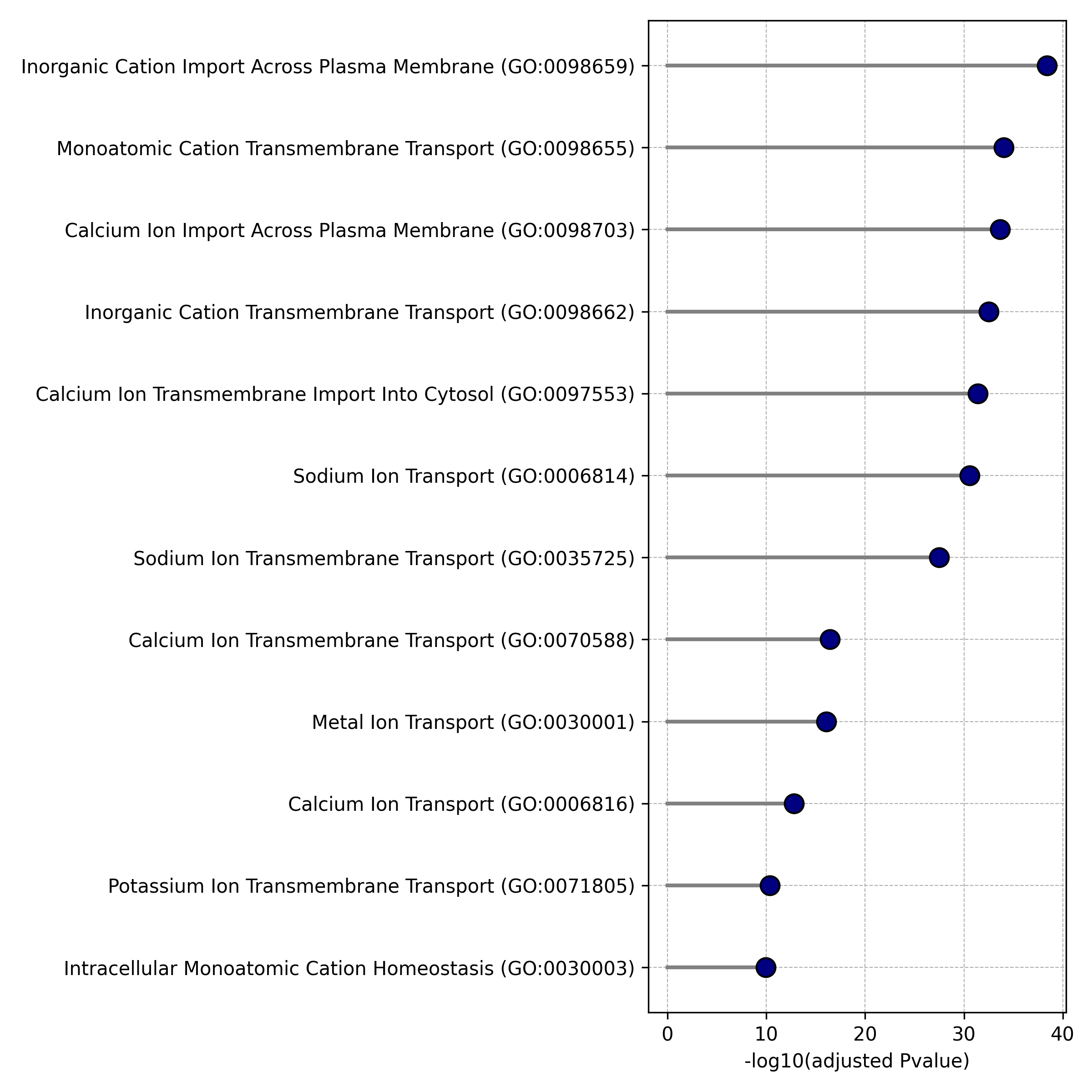}
    \end{minipage}
    \caption{\footnotesize Pahtway enrichment analysis for DN (left), HI (middle), and GOF (right) proteins.}
    \label{DN_HI_GOF_pathways}
\end{figure}

Pathway enrichment analysis revealed that DN proteins are enriched in pathways associated with filament organization (Figure~\ref{DN_HI_GOF_pathways}, left), HI proteins are overrepresented in pathways related to transcription regulation (Figure~\ref{DN_HI_GOF_pathways}, middle), and GOF proteins are enriched in pathways related to ion transport across membranes (Figure~\ref{DN_HI_GOF_pathways}, right).

\section*{DISCUSSION}

In this work, we introduce a novel framework that integrates GNNs with structural interactomics to predict both the MOI and the functional effects of mutated proteins in genetic disorders. By leveraging PPI network and high-resolution protein structures, we offer a graph-of-graphs approach that addresses two critical aspects of genetic disease prediction. This allows us to not only classify proteins as AD or AR but also predict whether AD diseases manifest through HI, GOF, or DN mechanisms.  Our framework demonstrated good performance in predicting MOI, with the GAT model achieving the best $F_1$ score for identifying AD and AR proteins. In terms of functional effects, the GCN model effectively classified HI, GOF, and DN proteins based on structural features.

The most important feature in predicting AD proteins is the evolutionary pressure score (UNEECON) \citep{Huang2020-pd}. This aligns with previous studies showing that AD proteins experience stronger negative selection than AR proteins \citep{Blekhman2008-tg,Rapaport2021-xx}. Additionally, we observed a strong enrichment of AD proteins in pathways related to gene expression regulation. Prior research has shown that transcription factors (TFs) are often dosage-sensitive, particularly haploinsufficient, leading to dominant disease phenotypes \citep{Seidman2002-cf,Van_der_Lee2020-ym}. This is consistent with the fact that many human birth defects and neurodevelopmental disorders are caused by mutations in a single copy of TFs and chromatin regulator genes \citep{Zug2022-qu}.

For AR proteins, we found that the pLI (probability of loss-of-function intolerance) index is  the most important feature. The pLI index estimates the likelihood that knocking out one copy of a gene will result in a phenotype \citep{exac2016}. Low-pLI genes typically exhibit functional redundancy or possess sufficient reserve capacity, allowing heterozygous carriers to remain asymptomatic \citep{Huang2010-xp}. We also observed that AR proteins are enriched in mitochondrial pathways, consistent with previous findings that the vast majority of nuclear-encoded mitochondrial disease genes follow a recessive inheritance pattern \citep{Gusic2021-ln}. This bias toward recessive inheritance likely arises because defects in energy metabolism generally become pathogenic only when both alleles are disrupted. As long as one allele remains functional, mitochondrial pathways can sustain baseline energy production, preventing deleterious consequences \citep{Kacser1981-ne,Deutschbauer2005-oa}.

Feature attribution analysis revealed that DN proteins are strongly associated with high RNA-binding scores, consistent with the previous observation that DN mutations are enriched in nucleic acid-binding pathways \citep{Badonyi2024}. One possible explanation is that DN mutations often occur at critical interaction sites, such as DNA/RNA-binding interfaces, where they allow the mutant protein to retain its ability to bind partners but disrupt the overall function of the interacting complex. Additionally, we observed an enrichment of DN proteins in pathways related to filament organization. This is likely due to the inherent susceptibility of filamentous and polymeric assemblies to "poisoning" by mutant subunits, which can incorporate into multimers and destabilize the entire structure. A well-documented example is keratin-related disorders, where keratins (type I and II intermediate filament proteins) form an essential cytoskeletal network in epithelial cells. Mutations in keratin genes lead to cell fragility and are inherited in an AD manner, with the mutations exerting their effect through a DN mechanism \citep{Smith2003-qk}.

A depletion in topological domain emerged as the most important feature in predicting HI proteins. According to UniProt, the topological domain annotation defines the subcellular compartment in which each non-membrane region of a membrane-spanning protein is located. Our findings indicate that DN and GOF proteins have a higher fraction of topological domain annotations compared to HI proteins, suggesting that HI genes encode fewer membrane-spanning proteins than DN and GOF genes. This distinction may reflect fundamental differences in the functional roles of these proteins and their sensitivity to dosage effects. Furthermore, HI proteins are significantly enriched in pathways related to transcriptional regulation, consistent with previous findings that transcription factors are frequently dosage-sensitive and particularly prone to haploinsufficiency \citep{Seidman2002-cf,Van_der_Lee2020-ym}. 
 
Finally, the most important feature for predicting GOF proteins is helix structure. This finding is consistent with previous reports showing that GOF variants are significantly more likely to occur in alpha helices \citep{Stein2023-dh}. We also observed that GOF genes are enriched in pathways related to ion transport across membranes, further supporting the structural-functional link between helices and membrane proteins. A notable example are epilepsy-associated genes: approximately 25\% of them encode ion channels, many of which causing epilepsy through a GOF mechanism \citep{Oyrer2018-kz}. This association is biologically plausible, as many membrane proteins have a core architecture of transmembrane helices, which are critical for gating and transport functions \citep{Sansom2000-dl,Fernandez-Quintero2021-ho}. 

Although our approach provides a comprehensive view of inheritance patterns and functional effects, it has several limitations. First, the availability of high-quality structural data for all human proteins remains limited, potentially affecting prediction accuracy \citep{Bertoline2023-er}. To ensure uniform coverage, we relied on AlphaFold-predicted structures, which offer high accuracy for well-folded domains but have notable drawbacks. Unlike experimentally resolved structures, AlphaFold does not capture conformational flexibility, ligand interactions, or post-translational modifications, all of which are critical for functional interpretation. Additionally, it is less reliable for intrinsically disordered regions and dynamic protein states, where structural plasticity plays a key role \citep{Perrakis2021-vj}. Beyond structural considerations, our reliance on existing PPI network data introduces potential biases, as interaction coverage varies across tissues and biological contexts \citep{Ziv2022-ii}. Furthermore, class imbalance in labeled training data may affect model performance, particularly for underrepresented functional categories. Finally, while our method effectively predicts the functional effects of AD proteins, it does not extend to other inheritance patterns or interactions influenced by polygenic or epistatic effects \citep{Phillips2008-dn}.

Moving forward, there are several avenues for expanding this work. Incorporating tissue-specific PPI networks and expression data could improve the precision of our predictions, especially for proteins with context-dependent functions \citep{Ziv2022-ii}. Additionally, expanding the model to account for more complex inheritance patterns, such as polygenic traits and epistasis, could provide a more comprehensive understanding of genetic disease \citep{Boyle2017-dk}. Moreover, improving the interpretability of models in biological contexts remains essential to derive more actionable insights from the predictions \citep{Chen2024-du}. Finally, integrating these predictions with other computational tools and databases could further enhance our understanding of genetic diseases by providing a more holistic view of their underlying mechanisms \citep{saadat-fellay-2024-dna,Saadat2025-sh,saadat2024finetuningesm2proteinlanguage,saadat2025exploring,ruiz2025benchmarking}.


\section*{RESOURCE AVAILABILITY}


\subsection*{Lead contact}


Requests for further information and resources should be directed to and will be fulfilled by the lead contact, Jacques Fellay (jacques.fellay@epfl.ch).

\subsection*{Materials availability}


This study did not generate new materials.

\subsection*{Data and code availability}


The data and code for this study is available \href{https://github.com/AliSaadatV/Structural-Interactomics/}{here}. 

\section*{ACKNOWLEDGMENTS}


This work was funded by the Swiss National Science Foundation via grant \#197721 and by the Swiss State Secretariat for Education, Research and Innovation via contribution to project "UNDINE", SBFI No. 23.00322.

\section*{AUTHOR CONTRIBUTIONS}


Conceptualization, A.S.; methodology, A.S..; investigation, A.S., and J.F.; writing-–original draft, A.S.; writing-–review \& editing, A.S., and J.F.; funding acquisition, J.F.; supervision, J.F.

\section*{DECLARATION OF INTERESTS}


The authors declare no competing interests.

\section*{DECLARATION OF GENERATIVE AI AND AI-ASSISTED TECHNOLOGIES}


During the preparation of this work, the authors used GPT-4 in order to improve writing and readability. 

\section*{SUPPLEMENTAL INFORMATION INDEX}




\begin{description}
  \item Table S1. The descriptions of protein features.
  \item Table S2. The descriptions of amino acid features.
  \item Table S3. Hyperparameter tuning results for the MOI prediction model.
  \item Table S4. Hyperparameter tuning results for the functional effect prediction model.
  \item Table S5. Prediction of MOI for all autosomal proteins.
  \item Table S6. Prediction of functional effect for all AD-predicted proteins.
  \item Figure S1: Distribution of RNA-binding scores based on DRNApred.
  \item Figure S2: The distribution of the fraction of topological domain based on UniProt annotations.
  \item Figure S3: The distribution of the fraction of helical residues based on UniProt annotations.
  \item Figure S4: Number of AD, AR, and ADAR predicted based on the selected GAT model.
  \item Figure S5: probability of AD (pAD) vs probability of AR (pAR) for all autosomal proteins.
  \item Figure S6: Number of proteins predicted based on the selected GCN model. Prediction was performed on all AD and ADAR proteins.
  \item Figure S7: Number of proteins predicted based on the selected GCN model. ADAR proteins were excluded for this calculation.
  \item Figure S8: Number of proteins predicted based on the selected GCN model. Only ADAR proteins were included for this calculation.
\end{description}

\newpage


\bibliography{references}

\bigskip


\newpage

\section*{STAR METHODS}
\label{methods}

\subsection*{Data collection}

\paragraph{Mode of inheritance}

We collected the MOI data from the Gene Curation Coalition (GenCC) \citep{DiStefano2022} as well as the Online Mendelian Inheritance in Man (OMIM) \citep{Hamosh2002}. For GenCC records, we kept records with definitive, strong, or moderate gene-disease clinical validity. We focused on autosomal proteins, due to intrinsic differences in MOI for X chromosome proteins. Proteins were accordingly labeled as AD, AR, or ADAR (both dominant and recessive).

\paragraph{Molecular mechanism}

We collected the functional effect of AD proteins from \citet{Badonyi2024}. This is a curated set of AD proteins labeled with their known functional effects, including DN, GOF, and HI. 

\paragraph{PPI network}

To make a comprehensive PPI network, we combined the interaction from four resources: STRINGdb with interaction score $\geq$ 0.7 \citep{Szklarczyk2022}, BioGRID \citep{Oughtred2020}, the Human Reference Interactome (HuRI) \citep{Luck2020}, and \citet{Menche2015}, which resulted in a network with 17,248 nodes, and 375,494 edges.

\paragraph{Protein graph}

We downloaded the predicted structures of all human proteins from the AlphaFold database \citep{Varadi2023}. We then used Graphein \citep{jamasb2022graphein} to construct a residue graph for each protein based on its structure. In these residue graphs, nodes represent amino acids, and edges capture various interactions between them, including peptide bonds, aromatic interactions, hydrogen bonds, disulfide bonds, ionic interactions, aromatic-sulfur interactions, and cation-$\pi$ interactions. To account for long-range amino acid interactions, we included edges between amino acids that are spatially close (\textless 5 angstroms) but distant in the sequence (\textgreater 5 amino acids apart).  

\paragraph{Protein features}

We annotated all proteins with 97 features covering various aspects of protein characteristics, including structure and function, conservation and constraint, and expression and regulation. Using the training set, we excluded features with low variance (\textless 0.1) or high correlation (\textgreater 0.8), resulting in a final selection of 78 features. The complete list of initial and selected protein features is provided in Supplementary Table 1.  

\paragraph{Residue features}

For the residue graphs, we annotated the nodes (i.e. amino acids) with 132 features covering various aspects including structure and function, sequence, biochemical, and evolutionary charecteristics. Using the training set, we excluded features with low variance (\textless 0.01) or high correlation (\textgreater 0.8), resulting in a final selection of 73 features. The complete list of initial and selected residue features is provided in Supplementary Table 2. 

\subsection*{Model development}

\paragraph{Study design}

In this study, MOI is predicted by classifying nodes in a PPI network, while functional effect prediction is framed as a graph classification task. Both models use a multi-label classification approach, allowing inputs to have more than one label. All experiments were conducted using the PyTorch Geometric library \citep{fey2019}.

We leverage GNNs because they can directly utilize the relational structure of the data by propagating information through the graph \citep{Khemani2024-le}, whereas traditional machine learning models require explicit feature engineering to incorporate graph-based information. For MOI prediction, PPI networks encode valuable topological properties that influence disease mechanisms. GNNs capture both individual protein features and their connectivity within the network, an aspect that conventional ML models struggle to integrate effectively. Similarly, for functional effect prediction, protein structure graphs provide spatial and biochemical context at the residue level. While traditional ML approaches require manually extracting graph-based features (e.g., graph centrality, residue connectivity), GNNs inherently integrate these properties, enabling a more comprehensive representation of structural and functional information.

\paragraph{Architecture}

For both MOI and functional effect prediction, we utilized various graph neural network architecture including graph convolutional network (GCN) \citep{kipf2017}, graph attention network (GAT) \citep{GATv2}, and graph isomorphism network (GIN) \citep{xu2019}. 

GCNs extend the concept of convolution from grid-like data (such as images) to graph data, allowing the aggregation of feature information from neighboring nodes. This approach effectively captures local graph structure and node features. The forward propagation formula in a GCN is given by:

\[
h_i^{(l+1)} = \sum_{j \in {\mathcal{N}}(i)} \frac{1}{\sqrt{\text{deg}(i)} \sqrt{\text{deg}(j)}} \mathbf{W}^{(l)} h_j^{(l)}
\]

\begin{itemize}
    \item \( h_j^{(l)} \): The node feature vector at layer \( l \). 
    \item \( h_i^{(l+1)} \): The updated node feature vector at layer \( l+1 \).
    \item \( \mathbf{W}^{(l)} \): The learnable weight matrix for layer \( l \).
    \item \( \mathcal{N}(i) \): The set of neighbors of node \( i \) (including itself due to the self-loop).
    \item \( \frac{1}{\sqrt{\text{deg}(i)} \sqrt{\text{deg}(j)}} \): The normalization term based on the degrees of nodes \( i \) and \( j \), ensuring that nodes with different degrees contribute proportionally to the update.
\end{itemize}

GINs are designed to be powerful for graph isomorphism, making them capable of distinguishing a wide variety of graph structures. They achieve this by using a multi-layer perceptron (MLP) to aggregate node features, enhancing their discriminative power. The update rule for the GIN is given by:

\[
h_i^{(l+1)} = \text{MLP}^{(l)}\left( \left(1 + \epsilon^{(l)} \right) h_i^{(l)} + \sum_{j \in \mathcal{N}(i)} h_j^{(l)} \right)
\]

\begin{itemize}

    \item \( h_i^{(l)} \): The node feature vector at layer \( l \).

    \item \( h_i^{(l+1)} \): The updated node feature vector at layer \( l+1 \). 
    
    \item \( \text{MLP}^{(l)} \): A multi-layer perceptron applied at layer \( l \), which acts as a learnable transformation function on the aggregated node features.
    
    \item \( \epsilon^{(l)} \): A learnable parameter at layer \( l \) that adjusts the contribution of the central node's own features \( h_i^{(l)} \).
    
    \item \( \mathcal{N}(i) \): The set of neighbors of node \( i \). The sum \( \sum_{j \in \mathcal{N}(i)} h_j^{(l)} \) aggregates the features of all neighbor nodes in layer \( l \).
 
\end{itemize}

GATs introduce attention mechanisms to GNNs, enabling nodes to assign different importance weights to their neighbors. This allows for more flexible and expressive feature aggregation, potentially improving performance on tasks where certain neighbors have more influence than others. The forward propagation rule for GAT is given by:

\[
h_i^{(l+1)} = \sigma\left( \sum_{j \in \mathcal{N}(i)} \alpha_{ij}^{(l)} \mathbf{W}^{(l)} h_j^{(l)} \right)
\]

\[
\alpha_{ij}^{(l)} = \frac{\exp\left( \text{LeakyReLU}\left( a^T \left[ \mathbf{W}^{(l)} ( h_i^{(l)} \| h_j^{(l)}) \right] \right)\right)}{\sum_{k \in \mathcal{N}(i)} \exp\left( \text{LeakyReLU}\left( a^T \left[ \mathbf{W}^{(l)} (h_i^{(l)} \| h_k^{(l)}) \right]\right)\right)}
\]

\begin{itemize}
    \item $h_i^{(l)}$: The node feature vector at layer \( l \).
    \item $h_i^{(l+1)}$: The updated node feature vector at layer \( l+1 \). 
    \item $\alpha_{ij}^{(l)}$: The attention coefficient between nodes $i$ and $j$.
    \item $\mathbf{W}^{(l)}$: The weight matrix at layer $l$.
    \item $a$: The learnable attention vector.
    \item $\|$: The concatenation operator.
    \item $\mathcal{N}(i)$: The set of neighbors of node \( i \).
    \item \( \sigma(\cdot) \): A non-linear activation function (ReLU in our implementation).

\end{itemize}

\paragraph{Hyperparameters}

In all models, we employed a single hidden layer, with the output layer comprising two units for MOI prediction (AD and AR) and three units for functional effect prediction (DN, HI, and GOF). To identify optimal configurations, for each model we evaluated 25 combinations of hyperparameters, varying the hidden layer size across five values (128, 64, 32, 16, and 8) and the learning rate across five values (ranging from \(10^{-2}\) to \(5 \times 10^{-4}\)). The results of hyperparameter tuning for MOI and functional effect prediction are available in Supplementary Tables 3 and 4, respectively.

Other hyperparameters were set to commonly used values, including a weight decay of \(5 \times 10^{-4}\) and a dropout rate of 0.3. For optimization, we employed the Adam optimizer \citep{Kingma2014-mq} with an adaptive learning rate scheduler (ReduceLROnPlateau), which dynamically adjusted the learning rate based on validation loss.

\paragraph{Training and evaluation}
\label{training_and_evaluation}

To create train, validation, and test splits, we clustered protein sequences using MMseqs2 \cite{Steinegger2017} with thresholds of 20\% coverage and 20\% sequence identity. For each model, the proteins were divided into 80\% training, 10\% validation, and 10\% testing sets, ensuring minimal data leakage by performing the splits at the cluster level. 

We trained each model using a binary cross entropy loss for maximum 100 epochs and used early stopping based on validation loss to avoid over-fitting. We evaluated each selected model on the unseen test data using $F_1$, precision, and recall scores. 

We benchmarked the performance of our model against previous state-of-the-art approaches. For MOI prediction, we compared our model with DOMINO \citep{Quinodoz2017-sc}, which predicts the probability of a protein's association with dominant disorders (pAD). We used our MOI test set and excluded any proteins present in DOMINO’s training data. Since no threshold was provided, we classified proteins as AD if pAD $>$ 0.6, AR if pAD $<$ 0.4, and ADAR otherwise.

For functional effect prediction, we compared our model with those from \citet{Badonyi2024}, which include three separate SVM classifiers: DN vs. LOF, GOF vs. LOF, and LOF vs. non-LOF. To evaluate performance in a multi-label classification setting, we combined the test sets from these models and utilized the pre-calculated probabilities. We did not benchmark against other traditional ML models, as \citet{Badonyi2024} had already compared multiple approaches including SVM, LightGBM, Random Forest, Logistic Regression, and MLP, and identified SVM as the best performer. Based on these findings, we focused our comparisons on the strongest traditional baseline.

\paragraph{Model explanation}
\label{explanation}

To study the importance of features, we utilized Integrated Gradients \citep{sundararajan2017} using Captum \citep{kokhlikyan2020}. Since this method works per sample, we applied it on correctly predicted samples in the test sets. We included samples with only one label for further interpretability. Finally, we averaged feature attributions across selected samples, and scaled them by dividing to the maximum attribution.

\subsection*{Proteome-wide inference}

\paragraph{MOI and molecular mechanism inference}

After selecting the final models for MOI and functional effect prediction, we predicted the MOI for all proteins in the PPI network (Supplementary Table 5). Afterwards, we predicted the functional effect for the subset of proteins that were predicted as AD or ADAR (Supplementary Table 6).

\paragraph{Enrichment analysis}

To study further the predictions, we used GSEApy \citep{Fang2022} to perform enrichment analysis \citep{Khatri2012}, which is a statistical method used to determine whether known biological functions or processes are over-represented in a protein list of interest (e.g. AD proteins). In this method, the enrichment significance is calculated based on the hypergeometric distribution, where p-value is the cumulative probability of observing at least $k$ proteins of interest annotated to a specific protein set. The formula for the p-value is given by:

\[
p = 1 - \sum_{i=0}^{k-1} \frac{\binom{M}{i} \binom{N-M}{n-i}}{\binom{N}{n}},
\]

where $N$ is the total number of proteins in the background distribution, $M$ is the number of proteins in that distribution annotated to the gene set of interest, $n$ is the size of the list of proteins of interest, and $k$ is the number of proteins in that list which are annotated to the gene set. We focused on pathways containing at least 10 and at most 500 genes to exclude pathways that are either too specific or too general. As a reference database, we used Gene Ontology (Biological Processes) \citep{Ashburner2000, GO2023} to understand functional landscape of proteins.

\end{document}